\documentclass[prb,superscriptaddress,amsmath,amssymb,twocolumn,floatfix]{revtex4}
\usepackage{natbib}
\usepackage{graphicx}         
\usepackage{dcolumn}          
\usepackage{bm}               
\usepackage{upgreek}
\usepackage{booktabs} 				
\usepackage{tabularx}
\usepackage{array}
\usepackage[colorlinks=true,urlcolor=blue,breaklinks=true]{hyperref}

\begin{document}

\title{Fully in-situ InAs nanowire Josephson junctions by selective-area growth and shadow evaporation}

\author{Pujitha Perla}
\affiliation{Peter Gr\"unberg Institut (PGI-9), Forschungszentrum J\"ulich, 52425 J\"ulich, Germany}
\affiliation{JARA-Fundamentals of Future Information Technology, J\"ulich-Aachen Research Alliance, Forschungszentrum J\"ulich and RWTH Aachen University, Germany}

\author{H. Aruni Fonseka}
\affiliation{Department of Physics, University of Warwick, Coventry
CV4 7AL, UK}

\author{Patrick Zellekens}
\affiliation{Peter Gr\"unberg Institut (PGI-9), Forschungszentrum J\"ulich, 52425 J\"ulich, Germany}
\affiliation{JARA-Fundamentals of Future Information Technology, J\"ulich-Aachen Research Alliance, Forschungszentrum J\"ulich and RWTH Aachen University, Germany}

\author{Russell Deacon}
\affiliation{RIKEN Center for Emergent Matter Science and Advanced Device Laboratory, 351-0198 Saitama, Japan}

\author{Yisong Han}
\affiliation{Department of Physics, University of Warwick, Coventry
CV4 7AL, UK}

\author{Jonas K\"olzer}
\affiliation{Peter Gr\"unberg Institut (PGI-9), Forschungszentrum J\"ulich, 52425 J\"ulich, Germany}
\affiliation{JARA-Fundamentals of Future Information Technology, J\"ulich-Aachen Research Alliance, Forschungszentrum J\"ulich and RWTH Aachen University, Germany}

\author{Timm M\"orstedt}
\affiliation{Peter Gr\"unberg Institut (PGI-9), Forschungszentrum J\"ulich, 52425 J\"ulich, Germany}
\affiliation{JARA-Fundamentals of Future Information Technology, J\"ulich-Aachen Research Alliance, Forschungszentrum J\"ulich and RWTH Aachen University, Germany}

\author{Benjamin Bennemann}
\affiliation{Peter Gr\"unberg Institut (PGI-9), Forschungszentrum J\"ulich, 52425 J\"ulich, Germany}
\affiliation{JARA-Fundamentals of Future Information Technology, J\"ulich-Aachen Research Alliance, Forschungszentrum J\"ulich and RWTH Aachen University, Germany}

\author{Koji Ishibashi}
\affiliation{RIKEN Center for Emergent Matter Science and Advanced Device Laboratory, 351-0198 Saitama, Japan}

\author{Detlev Gr\"utzmacher}
\affiliation{Peter Gr\"unberg Institut (PGI-9), Forschungszentrum J\"ulich, 52425 J\"ulich, Germany}
\affiliation{Peter Gr\"unberg Institut (PGI-10), Forschungszentrum J\"ulich, 52425 J\"ulich, Germany}
\affiliation{JARA-Fundamentals of Future Information Technology, J\"ulich-Aachen Research Alliance, Forschungszentrum J\"ulich and RWTH Aachen University, Germany}

\author{Ana M. Sanchez}
\affiliation{Department of Physics, University of Warwick, Coventry
CV4 7AL, UK}

\author{Mihail Ion Lepsa}
\affiliation{Peter Gr\"unberg Institut (PGI-10), Forschungszentrum J\"ulich, 52425 J\"ulich, Germany}
\affiliation{JARA-Fundamentals of Future Information Technology, J\"ulich-Aachen Research Alliance, Forschungszentrum J\"ulich and RWTH Aachen University, Germany}

\author{Thomas~Sch\"apers}
\email{th.schaepers@fz-juelich.de}
\affiliation{Peter Gr\"unberg Institut (PGI-9), Forschungszentrum J\"ulich, 52425 J\"ulich, Germany}
\affiliation{JARA-Fundamentals of Future Information Technology, J\"ulich-Aachen Research Alliance, Forschungszentrum J\"ulich and RWTH Aachen University, Germany}

\date{\today}

\begin{abstract}
Josephson junctions based on InAs semiconducting nanowires and Nb superconducting electrodes are fabricated in-situ by a special shadow evaporation scheme for the superconductor electrode. Compared to other metallic superconductors such as Al, Nb has the advantage of a larger superconducting gap which allows operation at higher temperatures and magnetic fields. Our junctions are fabricated by shadow evaporation of Nb on pairs of InAs nanowires grown selectively on two adjacent tilted  Si (111) facets and crossing each other at a small distance. The upper wire relative to the deposition source acts as a shadow mask determining the  gap of the superconducting electrodes on the lower nanowire. Electron microscopy measurements show that the fully in-situ fabrication method gives a clean InAs/Nb interface. A clear Josephson supercurrent is observed in the current-voltage characteristics, which can be controlled by a bottom gate. The excess current of 0.68 indicates a high junction transparency. Under microwave radiation, pronounced integer Shapiro steps are observed suggesting a sinusoidal current-phase relation. Owing to the large critical field of Nb, the Josephson supercurrent can be maintained to magnetic fields exceeding 1\,T. Our results show that in-situ prepared Nb/InAs nanowire contacts are very interesting candidates for superconducting quantum circuits requiring large magnetic fields.
\end{abstract}
\maketitle

\section{Introduction}

III-V semiconductor nanowires combined with superconducting electrodes are versatile building blocks for various applications in the field of quantum computation and experiments addressing fundamental aspects of quantum nanostructures. Josephson junctions formed by two superconducting electrodes bridged by a nanowire segment allows the control of the critical current by a gate voltage.\cite{Doh05,Xiang06,Guenel12} Such an approach, results in a much more compact superconducting circuit lay-out compared to the common flux-controlled one. This advantage is used e.g. in a gatemon qubit, a special form of the transmon qubit, in which the  Josephson junction in the qubit resonator circuit is controlled by a gate.\cite{deLange15,Larsen15,Luthi18,Kringhoj20} Furthermore, owing to the large Fermi wavelength of the electrons in the semiconductor combined with the small diameter of the nanowire, a finite number of discrete Andreev bound states that carry the Josephson supercurrent are formed. Coherent transitions between these discrete states can be used in Andreev level qubits for quantum circuits.\cite{Zazunov03,Woerkom17,Tosi19} Apart from these more conventional qubit applications, nanowire-superconductor hybrids are also very promising for topological qubits based on Majorana fermions.\cite{Mourik12,Deng12,Das12,Albrecht16,Zhang17}  

InAs and InSb are the common semiconductors of choice to form a highly transparent interface with the superconducting electrodes, a prerequisite for a sufficiently large supercurrent. Depending on the application different superconductor materials are deposited. Thus, for example, aluminium with a small superconducting gap and a small critical magnetic field but large superconducting coherence length is a common choice.\cite{Doh05,Das12,Guenel14} For higher temperatures or higher magnetic field operation, superconductors such as Nb and its alloys,\cite{Guenel12,Guel17,Zhang17,Carrad19} Pb,\cite{Paajaste15,Kanne20} or V, \cite{Spathis11,Bjergfelt19} are used. However, detailed analysis of these semiconductor/superconductor systems are just starting to emerge.

Until recently, the conventional methods to produce superconductor-semiconductor interfaces, are based on the semiconductor surface cleaning (by wet chemical etching or Ar$^+$ sputtering) prior to the  superconductor deposition.\cite{Guel17} The major issues with these ex-situ approaches are the presence of residual atoms and the semiconductor surface damage that results in a non-ideal interface. Consequently, a soft induced gap might form in the semiconductor nanowire with a significant density of states present within the superconducting gap induced by the proximity effect.\cite{Guel17} This effect is especially detrimental for topological qubits based on Majorana fermions. A successful method to circumvent these problems is the in-situ deposition of the superconducting material on the semiconductor nanowires.\cite{Krogstrup15,Guesgen17,Bjergfelt19,Carrad19} 

Another important issue is the residual material left on the structure from  wet-chemical etching technique normally used for the fabrication of Josephson junctions with a small gap. However, recently it was demonstrated that closely separated superconducting electrodes can also be achieved by shadow evaporation technique, i.e. either by using a nanowire which crosses another\cite{Gazibegovic17,Khan20} or using a patterned, suspended SiO$_2$ layer as a shadow or stencil mask.\cite{Bjergfelt19,Carrad19}

Here, we combine both of the above approaches to report the fabrication of a fully in-situ Josephson junction based on the evaporation of superconductor half-shells on InAs nanowires by a specially designed shadow evaporation technique. Pairs of NWs are selectively grown on different Si(111) adjacent facets in such a way that one NW shadows the other one situated at a small distance. Thus, the deposited junction width is determined by the nanowire diameter and the distance between them. Although our approach is applicable to almost any superconductor material, here we focused on Nb, since small separations between Nb superconducting electrodes are difficult to fabricate by other methodologies. The InAs/Nb interface of the Josephson junctions fabricated by our novel approach are analysed in depth by electron microscopy techniques. Low temperature transport properties of the Josephson junctions are also reported.

\section{Results and Discussion}

\begin{figure}[ht!]
    \centering
    \includegraphics[width=1.0\columnwidth]{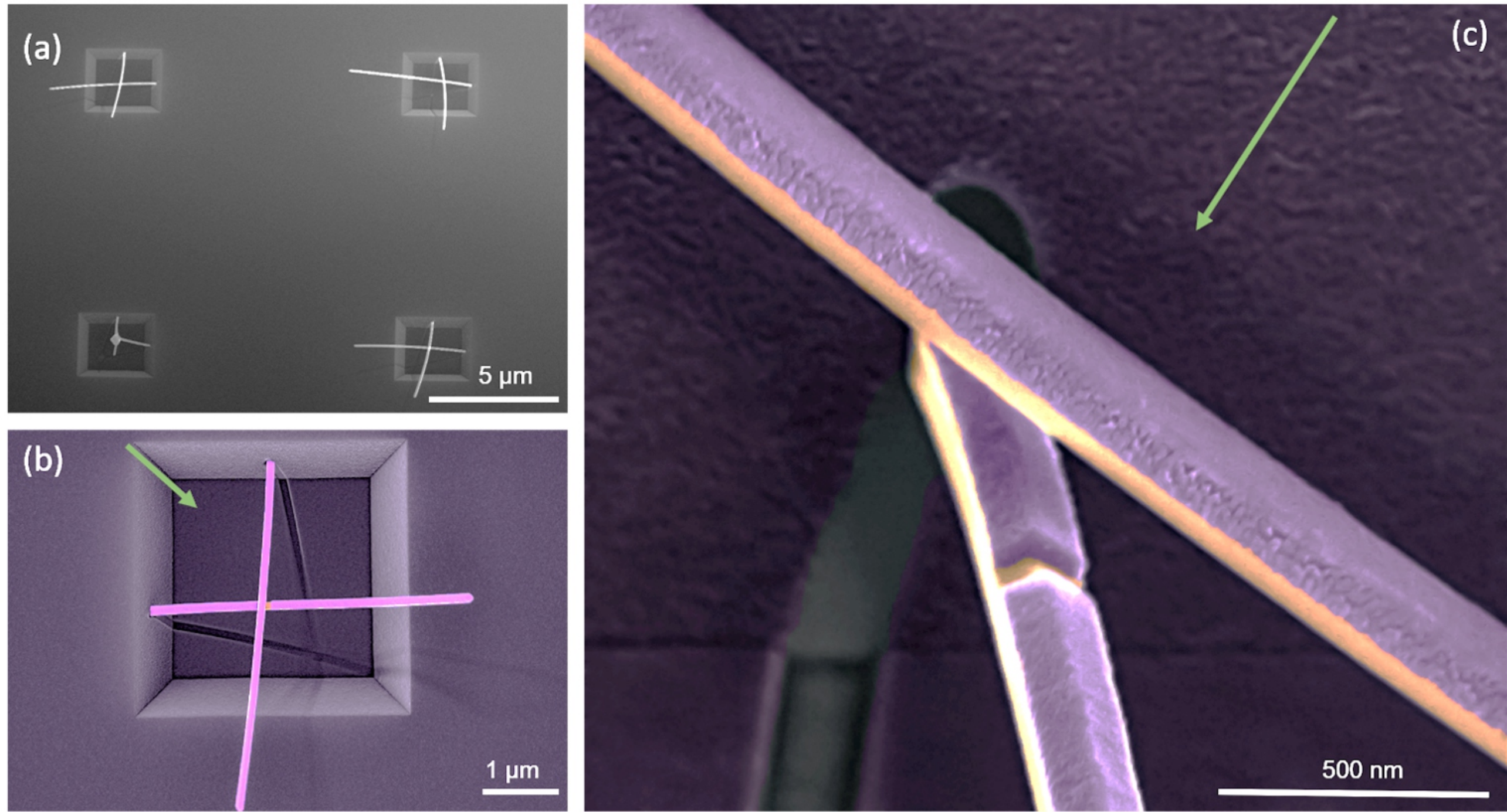}
    \caption{Scanning electron microscopy images of selectively grown nanowire junctions: (a) Overview of the 3\,$\mu$m square troughs with selectively grown nanowires (top view). (b)  False-coloured single square with two Nb covered nanowires grown off the adjacent Si (111) facets. The shadow depicts the direction of the metal (Nb) deposition (green arrow). (c) Close-up of the crossing section, showing the gap in the Nb layer (purple) in the bottom nanowire (orange) due to the shadowing from the top nanowire (30$^\circ$ titled image). }
    \label{fig:SEM-selective}
\end{figure}

\textbf{In-situ prepared InAs-Nb junctions.}
InAs-nanowire growth was carried out by molecular beam epitaxy (MBE). To enable selective area growth, 3\,$\mu$m wide square-shaped troughs are etched on to SiO$_2$-covered Si(100) substrates to obtain Si (111) side facets. Subsequently, nano-holes are etched on the side facets, defining the position of the nanowires. An offset of 100\,nm from the center of the facet is imposed to enable nanowires from neighboring facets to cross each other closely without merging. Following the growth of $4-5\,\mu$m long and 80\,nm diameter  InAs nanowires, the sample was transferred to a metal MBE chamber for the Nb half-shell deposition. A scanning electron microscopy (SEM) image with crossed nanowires are shown in Figure~\ref{fig:SEM-selective}a. A false-coloured magnified top-view image of one square trough is presented in Figure~\ref{fig:SEM-selective}b with two Nb-covered nanowires grown on adjacent Si (111) facets. The Nb gap on the bottom nanowire is clearly observed in the further magnified image in Figure~\ref{fig:SEM-selective}c, which confirms the formation of the Josephson junction with the separation of the two Nb electrodes defined by the shadow of the upper nanowire.
 
\begin{figure*}
    \centering
    \includegraphics[width=0.80\textwidth]{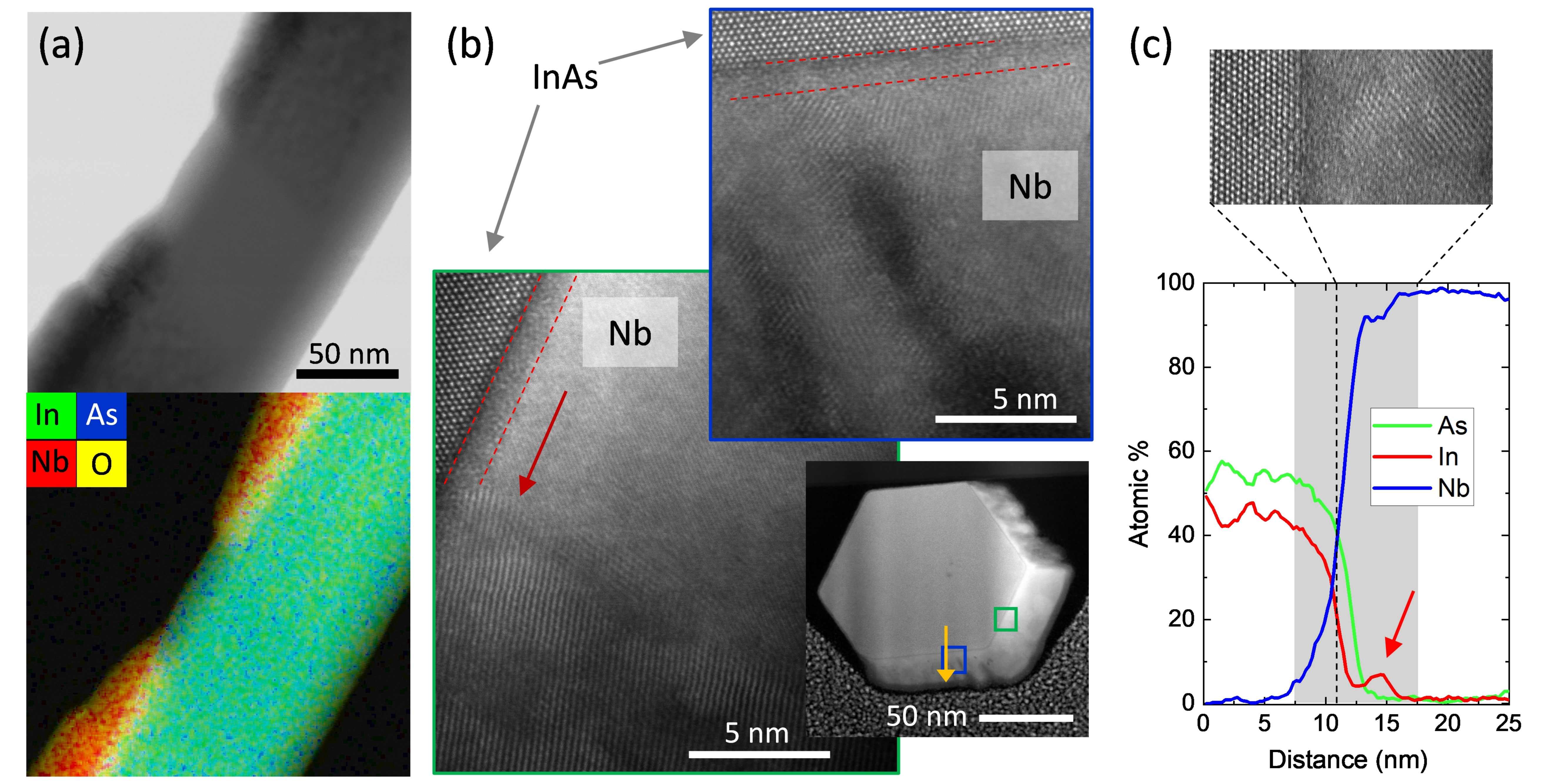}
    \caption{(a) Bright field and annular dark field (ADF) STEM images of a junction, with the EDX elemental maps superimposed on the ADF image. Both images show the clean gap in Nb. (b) ADF image of a nanowire cross section and higher magnification images from the regions indicated by squares. The amorphous layers are marked by the red broken lines in the higher magnification images. The red arrow points to a grain boundary. (c) EDX line scan profile along the yellow arrow in (b). The red arrow indicates the increase in In within the Nb layer and the greyed area marks the interface region shown in the inset high magnification ADF image.}
    \label{fig:TEM-Nb-InAs}
\end{figure*}

\textbf{Transmission Electron Microscopy.}
Due to the shadowing from the closely placed thin upper nanowire, junctions with separations of few tens of nanometers can be formed.  Figure~\ref{fig:TEM-Nb-InAs}a shows a bright field scanning transmission electron microscope (STEM) image of a Nb-InAs junction and the corresponding energy dispersive x-ray (EDX) elemental map, superimposed on the annular dark field (ADF) image. The separation is $\sim 55$\,nm and the junction is clean, i.e. with no traces of Nb within the gap. Along rest of the nanowire, a uniform and continuous Nb layer is formed, with only $\pm 2$\,nm variation in thickness along the nanowire. The side-view TEM studies also revealed that the nanowire has a polytypic crystal structure with thin wurtzite (WZ) and zincblende (ZB) segments, and high density of stacking faults within the segments (see Figure~S5a in Supporting Information (SI)).

Figure~\ref{fig:TEM-Nb-InAs} (b) shows an ADF image of a nanowire cross-section, along with two higher magnification images of the interface from the regions indicated. Nb is deposited on three of the six (\{11$\bar{2}$0\} type) side facets, in agreement with the SEM observations in Figure~\ref{fig:SEM-selective}c. Deposition on the middle facet is thicker ($\sim$22 nm), smooth and formed by relatively large grains of $\sim15-30$\,nm (one grain boundary is indicated by a red arrow). In contrast, those on the two facets on the sides are column-like, polycrystalline in structure and lower in thickness ($\sim$16 nm). This is due to the difference between the effective deposition angles on different facets. The metal flux is almost perpendicular to the nanowire axis at 87$^\circ$, and is directed at the middle facet. Which, also aided by the substrate temperature,\cite{Guesgen17} results in a smooth growth on this facet. The effective angles created with the two facets on either side are steep with a smaller deposition angles, resulting in column-like growth.\cite{Tanvir08}

Closer inspection of the InAs-Nb interface reveals a $\sim$1\,nm uniform amorphous layer on all three facets (marked by the red dashed lines). Figure~\ref{fig:TEM-Nb-InAs}c shows an EDX line scan across the interface at the position indicated by the yellow arrow in Figure~\ref{fig:TEM-Nb-InAs}b.  Two interesting observations can be made. Firstly, the amorphous region contains a high percentage of As with a clear lag visible in the decrease of the As curve compared to that of In. Secondly, In, which initially shows a dip within the amorphous region, slightly increases afterwards, before decreasing again (red arrow). The compositions of the amorphous layer measured across different facets and nanowires were found to vary between As:\,25-40\%\, In:\,5-20\%\ and, Nb:\,45-60\%\ (considering only Nb, As and In). One example is shown in SI Figure~S5b. Although the composition values of the 1 nm layer cannot be ascertained precisely due to the contributions from layers on either side, it is clear that this region contains a higher percentage of As and Nb. Considering the ternary phase diagram between In-As and Nb at room temperature,\cite{Klingbeil89} one could see that there is no tie-line between InAs and Nb. This means that InAs and Nb cannot exist in equilibrium. Instead, they react in a dominant reaction\cite{Klingbeil89} and form compounds, even at room temperature. The crystal structure of the layer formed by mixing (or solid diffusion) is amorphous, similar to many semiconductor-metal interfaces that show similar behaviour of solid state amorphisation.\cite{Zhang17,Chen20,Sinclair94}

As the amorphous layer contains much less In than As, the excess In from InAs decomposition is expelled and segregated on the far Nb side, forming an In rich band. The existence of tie-lines between number of Nb$_x$As$_y$ and In/Nb$_3$In in the In-As-Nb phase-diagram,\cite{Klingbeil89} suggests that compounds of the former can co-exist with In or Nb$_3$In. Similar observations have been made in other material systems such as Pt-GaAs, where gallide formation took place close to the metal interface and arsenide formation close to the semiconductor interface during re-crystallisation.\cite{Sinclair94} Although the subsequent transport measurements do not indicate significant effects from this amorphous layer, the current results bring to notice the important aspect of room temperature reactions and amorphisation of semiconductor-superconductor/metal interfaces.

\begin{figure*}[ht!]
    \centering
    \includegraphics[width=0.80\textwidth]{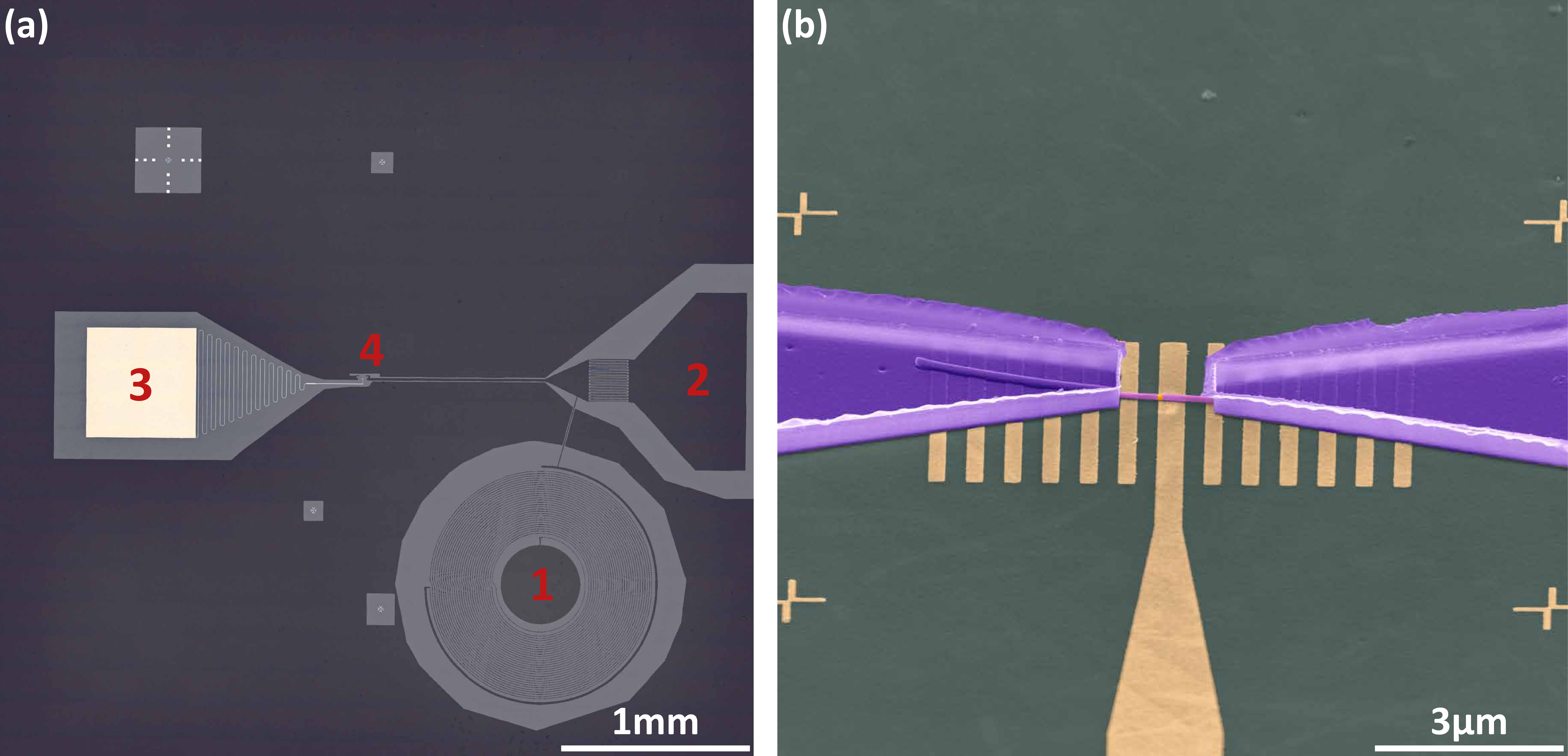}
    \caption{(a) Optical microscope image of a bias-tee chip implemented by combining a coil (1) and an inter-digital capacitor (2) connected to one side of the junction. The other side is connected to the global ground plane. For electrostatic tuning, we use a bottom gate electrode, which is terminated by a large bonding pad (3). The junction is located at (4). (b) Scanning electron micrograph of an InAs nanowire covered by Nb half-shells, which are contacted by NbTi fingers. The junction is placed on a bottom-gate electrode. The metal finger grid on either sides of the gate are for mechanical support of the nanowire.}
    \label{fig:device}
\end{figure*}

\begin{figure*}[!t]
    \centering
    \includegraphics[width=0.90\textwidth]{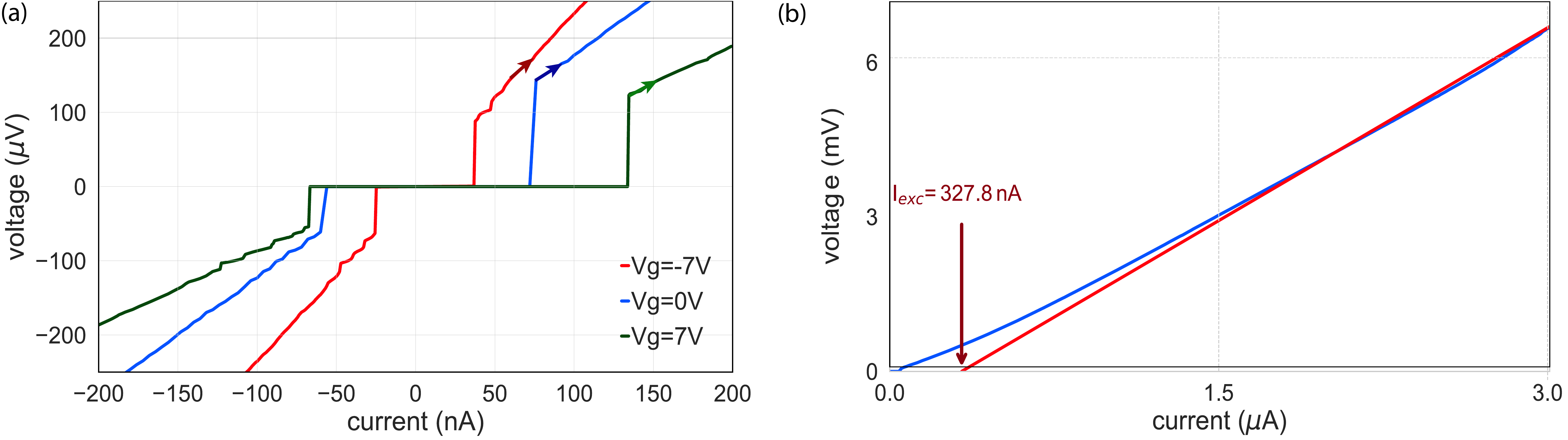}
    \caption{(a) Current-voltage characteristics of an InAs/Nb shadow junction measured at $V_{\mathrm{g}}=0$\,V ,  $-7$\,V, and $7$\,V, showing an almost three-fold reduction of the critical current between the largest and smallest gate voltages. The sweep direction is indicated by arrows. The junction is slightly underdamped and shows a small hysteretic behaviour due to overheating. (b) $IV$ characteristics of the same junction for large bias currents and a gate voltage of $V_{\mathrm{g}}$=7$\,$V. Based on the measurement we obtain an excess current $I_\mathrm{exc}=327.8\,$nA and a normal state resistance $R_\mathrm{N}=2850\,\Omega$.}
    \label{fig:IV}
\end{figure*}

\textbf{Electrical Characterization.}
 The nanowire Josephson junction was integrated in an on-chip bias tee, containing an inter-digital capacitor and a planar coil, to perform both AC and DC measurements (cf. Figure~\ref{fig:device}a).
The Nb shells were contacted by NbTi fingers, while control of the carrier concentration of the InAs segment between the Nb electrodes was achieved by means of a bottom gate. A scanning electron micrograph of the junction device is depicted in Figure~\ref{fig:device}b. In order to get an overview of the junction properties the current-voltage ($IV$) characteristics at three different gate voltages at the temperature of 15\,mK were measured. As can be seen in Figure~\ref{fig:IV}a, at zero gate voltage a relatively large switching current of $I_\mathrm{c}=75$\,nA and a re-trapping current of $I_\mathrm{r}=60$\,nA are observed, which is inline with previous studies on Nb/InAs nanowire junctions.\cite{Guenel12} For larger gate voltages, i.e. $V_\mathrm{g}=7$\,V, those values increase to $I_\mathrm{c}=133$\,nA and $I_\mathrm{r}=67$\,nA, respectively. Whereas, for $V_\mathrm{g}=-7$\,V the switching and retrapping currents are lowered to values of about 40\,nA and 25\,nA, respectively. Thus, between the smallest and largest gate voltage the junction exhibited an almost tripling of the switching current. The fact that $I_\mathrm{r}$ is only slightly lower than $I_\mathrm{c}$ indicates that the heating effect caused by dissipation in the resistive state is only moderate.
As a result of the special circuit geometry, namely the coupling to a coplanar waveguide transmission line, the junction is affected by the emission and self-absorption of photons due to the AC Josephson effect, resulting in so-called self-induced Shapiro steps on the re-trapping branch. The decrease $I_\mathrm{c}$ with decreasing $V_\mathrm{g}$ can be attributed to the fact that by lowering the carrier concentration the number of transport channels carrying supercurrent via phase-coherent Andreev reflections is reduced. However, for even more negative gate voltages, no complete suppression of $I_\mathrm{c}$ could be achieved. We attribute this to the incomplete pinch-off of the electron gas in the InAs nanowire bridge segment. This is in contrast to our Al-based nanowire junctions for which all transport could be completely suppressed.\cite{Zellenkens20a} A possible reason is that the in-situ deposited Nb layer changes the Fermi level pinning at the interface leading to an enhanced carrier accumulation at the interface. Furthermore, the structural properties, like the alloying and the amorphous layer at the interface may have an affect as well. As a consequence of the inability to pinch-off the junction, no tunnel spectroscopy could be performed, in order to find out about the hardness of the induced gap. However, due to the large superconducting gap of Nb, e.g. compared to the gap of Al, a more robust supercurrent is maintained in the junction. 
\begin{figure*}[!t]
    \centering
    \includegraphics[width=0.90\textwidth]{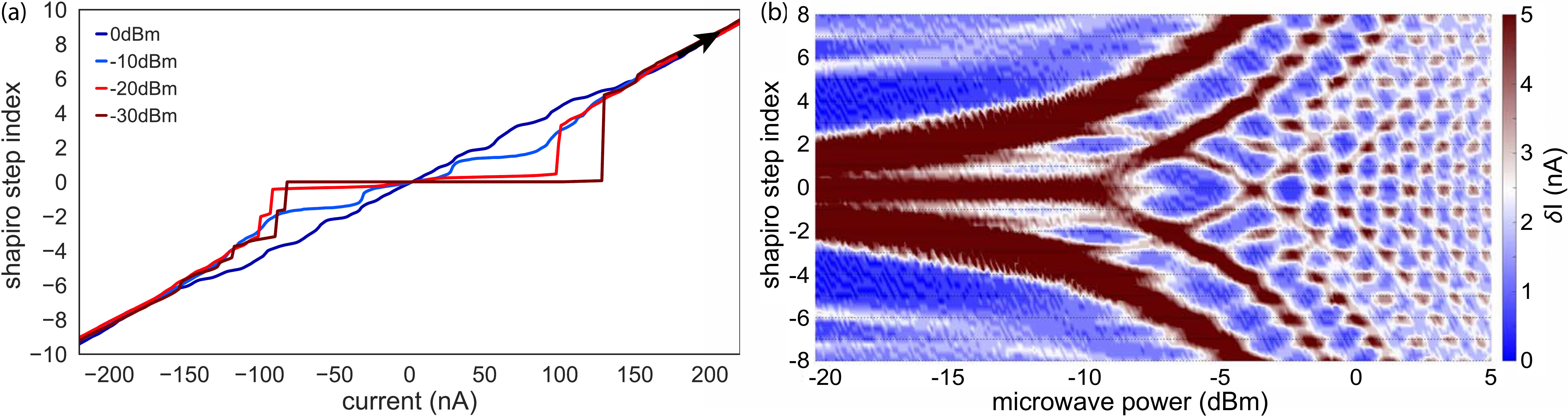}
    \caption{(a) Current-voltage traces for different microwave excitation powers at a fixed frequency of $f=5$\,GHz. While the trace at $-30\,$dBm (dark red) reproduces the current-voltage curves without any additional AC component, the pronounced plateau region within the zero voltage state is replaced by equidistant voltage steps when the power is increased. The sweep direction for all measurements is indicated with the black arrow. (b) Histogram of the power-dependent Shapiro response for a constant microwave frequency of $f=4\,$GHz.}
    \label{fig:Shapiro}
\end{figure*}

In order to obtain information about the junction transparency, we measured the $IV$ characteristics up to large bias voltages at $V_\mathrm{g}=0\,$V. 
By linear extrapolation in the bias voltage range above $2\Delta/e$ we were able to extract an excess current of $I_{\mathrm{exc}}=$327.8$\,$nA and a normal state resistance of $R_{\mathrm{N}}=$2850$\,\Omega$. By utilizing the framework of the corrected Octavio--Tinkham--Blonder--Klapwijk theory,\cite{Octavio83,Flensberg88} we obtain a ratio of $eI_\mathrm{exc}R_\mathrm{N}/\Delta=$0.623 which results in a barrier strength of $Z=0.69$. The latter corresponds to a transparency $\mathcal{T}=0.68$, which is a typical value for a nanowire Josephson junction with a wide-gap superconductor like Nb in the many channel regime.\cite{Guenel12} The junction transparency is large, with no significant detrimental effect apparent from the amorphous interfacial layer observed in the TEM studies.  

For mesoscopic Josephson junctions, the device response to an applied microwave signal provides information about the internal state structure, such as presence of a topological state.\cite{Dominguez17,Bocquillon17} Thanks to the combination of the coplanar waveguide transmission line and the bias-tee, the nanowire Josephson junction can be supplied with an AC and DC signal simultaneously with efficient transmission to the junction over a wide range of frequency, which would be difficult to achieve with an external antenna. Figure~\ref{fig:Shapiro}a shows a set of current-voltage traces for a fixed frequency of $f=5\,$GHz and different microwave powers at $V_{\mathrm{g}}=7\,$V. For low power, i. e. $-30\,$dBm, the curve mimics the behavior of a purely DC-driven junction. However, if the power is increased, the zero voltage state is gradually suppressed and replaced by equidistant voltage plateaus, so-called Shapiro steps, of height $n \times hf/2e$, with $h$ Plancks's constant and $n=1,2,3, \dots$. Originating from the AC Josephson effect, they are tightly bound to the current-phase relation and damping behavior. However, while all integer steps are well pronounced, indicating a single well-defined junction, there is no obvious indication for any non-trivial features such as missing steps. Most importantly, there are no signs of subharmonic steps, which may be observed as a consequence of a non-sinusoidal current-phase-relationship due to a  interface transparency close to unity.

As one can see in Figure~\ref{fig:Shapiro}a, the quality and shape of the Shapiro steps also depend on the effective microwave power that is applied to the junction. Thus, we performed a more systematic mapping of the AC response for constant frequencies. Figure~\ref{fig:Shapiro}b shows the histogram of binned voltage data scaled by the current step size of the measurement at $f=4\,$GHz for microwave powers at the input port of the fridge between $-20\,$dBm and $+5\,$dBm. For low powers, the junction exhibits a chevron-like pattern without well-defined steps. The latter can probably be attributed to the fact that the small amplitude of the AC drive is not sufficiency to maintain a resonant motion of the particle in the washboard potential and resistively shunted junction (RSJ) model. However, when the microwave power is increased above $-5\,$dBm, clearly pronounced Shapiro steps are observed. 
\begin{figure*}[!t]
    \centering
    \includegraphics[width=0.90\textwidth]{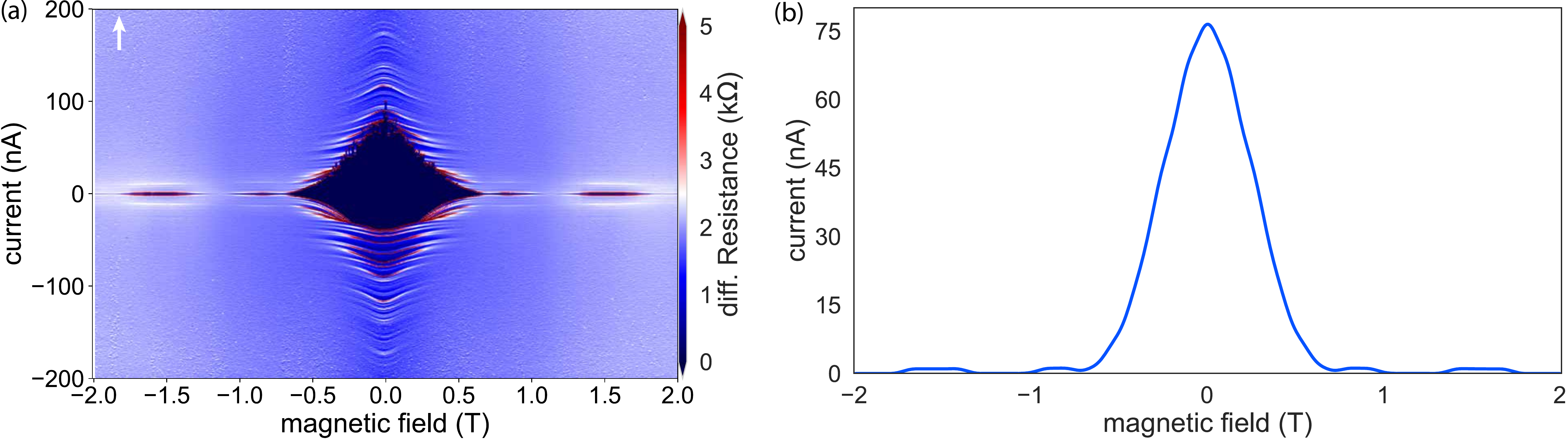}
    \caption{(a) Magnetic field dependent differential resistance for V$_g\,=\,$0$\,$V. The field is oriented in-plane along the nanowire axis. For both sweep directions, the nanowire junction exhibits a fluctuating resistance that corresponds to the alternating suppression and revival of the supercurrent. This effect can be attributed to a mixture of spin-orbit interaction and the interference between multiple transverse modes in the nanowire.\cite{Zuo17} The observed behavior is maintained for magnetic fields above 2$\,$T, indicating a comparably large critical field B$_c$. (b) Field-dependent magnitude of the switching current, clearly showing the reappearance of the supercurrent for fields up to 2$\,$T.}
    \label{fig:Fraunhofer}
\end{figure*}

Operation in the presence of a magnetic field is common to all structures that are based on few-channel mesoscopic nanowire Josephson junctions, i.e. the Andreev qubit or topological systems. In the case of the former, for example, the system can operate as intended if the junction and the connected superconducting loop is exposed to a magnetic flux $\Phi=\Phi_0/2$ corresponding to a phase bias of $\pi$, with magnetic flux quantum $\Phi_0=h/2e$. For the creation of Majorana zero modes, on the other hand, one needs a strong in-plane field that can easily exceed hundreds of milli-Tesla. This is especially true in the case of InAs due to the smaller g-factor if compared with InSb.\cite{Winkler03}. Thus, the magnetic field robustness of the induced superconductivity in the nanowire junction is of special interest to benchmark the device performance. Figure \ref{fig:Fraunhofer}a shows the device response in terms of the change of the differential resistance, if the system is penetrated by a magnetic field in parallel to the nanowire axis. Here, the most obvious feature is the lobe-like pattern centered around zero magnetic field. Considering the center lobe, the supercurrent is suppressed at large magnetic field magnitudes of around  $\pm 0.7$\,T. The strong asymmetry along the current axis close to zero magnetic field can be attributed to the difference of the switching and retrapping current. The stripe-like structure in the retrapping branch is due to the existence of self-induced Shapiro steps. We find that the field-dependency of the supercurrent does not follow the expected monotonous decrease when the superconducting gap energy decreases with the magnetic field. Instead, the device exhibits an alternating, non-periodic series of sections with and without a supercurrent. These lobe-like structures at higher magnetic fields, in which the supercurrent reappears for a finite field range, can probably be attributed to the intermixing and interference of multiple but not-too-many transverse modes.\cite{Zuo17} The corresponding, field-dependent magnitude of the switching current depicted in Figure~\ref{fig:Fraunhofer}b clearly shows that the lobe-structures do not follow a typical Fraunhofer-like pattern and the junction can still host finite supercurrent even up to 2\,T.

\section{Conclusion}

Our results show that highly transparent Josephson junctions can be fabricated by combining selective-area growth with a shadow evaporation scheme for the superconducting electrodes. The transmission electron microscopy investigations confirmed the absence of any foreign residue at the interface between the superconductor and the InAs nanowire. However, a very thin amorphous layer is observed at the interface. The Nb growth on the middle facet is found to be smooth consisting of large grains, while the side facets are polycrystalline and column-like. Owing to the large interface transparency, the junctions showed a clear signature of a Josephson supercurrent in the transport experiments. Gate control was possible, however, compared to Al-based junctions prepared in a similar fashion no complete pinch-off was achieved, which may be due to an enhanced surface accumulation in InAs when in contact to Nb. The Shapiro steps observed in the $IV$ characteristics show pronounced integer steps, indicating a sinusoidal current-phase relation. Taking the magnetic interference effects into account we have strongly evidenced successful fabrication of a weak link which works well even in high magnetic fields.

Our Nb/InAs nanowire-based junctions prove to be very interesting devices, with great potential for applications in superconducting quantum circuits that require large magnetic fields. In fact, most of the advanced approaches for the detection, manipulation and utilization of topological excitations, like the Majorana-transmon, rely on phase-sensitive and well-controlled detector structures. Here, our InAs/Nb in-situ nanowire shadow Josephson junctions can act as ideal gate-tunable components in superconducting quantum interference device (SQUID) structures which do not exhibit pronounced quantum fluctuations of the supercurrent and, additionally, maintain their superconducting properties even at large magnetic fields. The exact origin of the non-ideal transparency remains unclear. Even though we do not see any obvious indications in our measurements that the transport properties are altered by the interlayer, it still interesting for its effects to be investigated experimentally. Further improvement of the nanowire Josephson junctions may be possible by post-growth annealing and hence inducing re-crystallisation of the amorphous layer at the Nb/InAs interface.

\section{Methods}

\textbf{Growth and Fabrication.} 
The pre-pattered Si substrates used for the selective-area growth of the InAs nanowires were prepared by using a three-step electron beam lithography process, i.e. the first step places the alignment marker, the second defines the square-shaped troughs, and the third step is employed to define the nanoholes on the side facets of the square troughs. The substrates used for the template fabrication are Si (100) wafers, thermally oxidized for 20\,nm. The alignment markers for electron beam lithography are defined by deep etching using reactive ion processing. Next, several sets of 3\,$\mu$m wide square troughs with a pitch of 10\,$\mu$m are defined on the substrate surface (cf. Figure~S1a in Supporting Information), Here, a PMMA resist layer is used as a mask to anisotropically etch the SiO$_2$ layer by reactive ion etching using CHF$_3$ and oxygen revealing the Si(100) surface. Subsequently, the resist mask is removed and the Si(100) surface is etched for 90\,s with tetramethyl ammonium hyroxide (TMAH) to form the 300\,nm deep square-shaped troughs with the Si(111) facets (cf. Figure~S1b). Next, the oxide on the substrate surface is removed completely with buffered HF. Subsequently, a thermal re-oxidation is performed resulting in a 23 and 16\,nm thick oxide layer on the (111) and (100) surfaces, respectively (cf. Figure~S2). As a part of the third lithography step, 80\,nm wide growth holes are defined in the oxide layer on the side facets for the subsequent selective-area growth. The holes are etched using a combination of reactive ion etching and HF wet etching (cf. Figure~S1c). A focused ion beam etching prepared cross-sectional cut of a 80\,nm wide hole on a Si (111) facet is depicted in Figure~S1d. Regarding the position of the holes an offset of 100\,nm from the center of the facet is imposed to enable nanowires from neighboring facets to cross each other closely rather than merging into one crystal.

The InAs nanowires are selectively grown in the holes on the Si(111) facets via molecular beam epitaxy (MBE). A vapour-solid method without any catalyst is employed. In the first step the nanowires are grown at a substrate temperature of 480\,$^{\circ}$C with an indium growth rate of 0.08 $\mu$m/h and an As$_4$ beam equivalent pressure (BEP) of $\approx 4 \times 10^{-5}$\,mbar for 10\,min to sustain an optimal growth window and then in the second step the substrate temperature is decreased to 460\,$^{\circ}$C with an indium growth rate of 0.03\,$\mu$m/h and As$_3$ BEP of $\approx 3 \times 10^{-5}$\,mbar for 2.5\,h resulting in $4-5\,\mu$m long and 80\,nm wide nanowires.

After the growth of the InAs nanowire, the substrate undergoes an arsenic desorption at 400\,$^\circ$C for 20\,min and at 450\,$^\circ$C for 5\,min. Subsequently, the sample is transferred to a metal MBE chamber. Further on, the Nb metal shell is deposited at an angle of 87$^\circ$ to the nanowire axis. The Nb is evaporated at a substrate temperature of 50$\,^\circ$C. The measured growth rate of 0.082\,nm/s resulted in an average Nb thickness on the nanowire of 17\,nm.

Further processing details on substrate preparation and growth can be found in the Supporting Information. In addition, as elaborated in the Supporting Information, in-situ InAs nanowire-based Josephson junctions can also be fabricated by using random nanowire growth on adjacent Si(111) side facets. This approach offers the advantage of easier fabrication but has the disadvantage of uncontrolled junction formation (cf. Figure S4).

\noindent \textbf{Device fabrication.}
The devices for electrical characterization were fabricated on highly-resistive Si substrates with pre-patterned bottom gate structures and a superconducting circuit, as shown in Figure~\ref{fig:device}a. As the devices are intended to work for both AC and DC measurements, we use a  transmission line in coplanar waveguide geometry to form the source contact of the nanowire Josephson junction. The latter is terminated by an on-chip bias tee, consisting out of an inter-digital capacitor and a planar coil. All three elements, together with the surrounding ground plane, were made of reactively sputtered TiN with a thickness of 80$\,$nm. Subsequently, the nanowires were deposited onto the electrostatic gates by means of a SEM-based micro-manipulator setup. To ensure an ohmic coupling between the contacts, made out of NbTi, and the Nb shell, we used an in-situ  Ar$^+$ dry etching step prior to the metal deposition. The contact separation is chosen to be at least 1.5$\,\mu$m in order to reduce the effect of the wide-gap superconductor NbTi on the actual junction characteristics. The finished junction device is depicted in Figure~\ref{fig:device}b. \\
\noindent \textbf{Transmission electron microscopy}
 For the side view analysis, the nanowires were transferred from growth arrays to holey carbon grids by gently rubbing the two surfaces. The cross section samples were prepared using focused ion beam (FIB). TEM analysis was carried out using doubly corrected Jeol ARM 200F and Jeol 2100 microscopes, both operating at 200\,kV. The EDX measurements were carried out using an Oxford Instruments $100\,\mathrm{mm}^2$ windowless detector installed within the Jeol ARM 200F.

\noindent \textbf{Electrical measurements.} The electrical measurements were performed in a $^3$He/$^4$He dilution refrigerator with a base temperature of 13\,mK. The current-voltage characteristics were measured in a quasi four-terminal configuration using a current bias. For the differential resistance measurements a standard lock-in technique was employed. The rf-frequency signal for the measurements of the Shapiro steps was applied to the junction via the capacitor of the bias-tee.

\section*{Acknowledgements}
We  thank Tobias Ziegler and  Anton Faustmann for their helpful discussions and the assistance with the micro-manipulator, Michael Schleenvoigt with the metal deposition, Christoph Krause and Herbert Kertz for technical assistance. Dr. Florian Lentz and Dr. Stefan Trellenkamp are also gratefully acknowledged for electron beam lithography. Dr. Elmar Neumann and Stephany Bunte for their immense help with the FIB and the Magellan SEM assistance. Dr. Gianluigi Catelani is gratefully acknowledged for theory support regarding the magnetic field measurements. All samples have been prepared at the Helmholtz Nano Facility.\cite{GmbH2017} The work at RIKEN was partially supported by
Grant-in-Aid for Scientific Research (B) (No. 19H02548), Grants-in-Aid for Scientific Research (S) (No. 19H05610), and Scientific Research on Innovative Areas "Science of hybrid quantum systems" (No. 15H05867). The work at Forschungszentrum J\"ulich was partly supported by the project "Scalable solid state quantum computing", financed by the Initiative and Networking Fund of the Helmholtz Association. UK EPSRC is acknowledged for funding through grant No. EP/P000916/1. 


%

\renewcommand{\thesection}{S\Roman{section}}
\renewcommand{\thesubsection}{\Alph{subsection}}
\renewcommand{\theequation}{S\arabic{equation}}
\renewcommand{\thefigure}{S\arabic{figure}}
\renewcommand{\figurename}{Figure}
\setcounter{section}{0}
\setcounter{equation}{0}
\setcounter{figure}{0}
\setcounter{table}{0}
\setcounter{page}{1}
\makeatletter
\renewcommand{\bibnumfmt}[1]{[S#1]}
\renewcommand{\citenumfont}[1]{S#1}

\begin{center}
{\large \bf Supplementary Information}
\end{center}

\section{Sample fabrication and growth} \label{sec:fab-growth}

\subsection{Preparation of the patterned substrate}
\begin{figure}[h!]
	\centering
	\includegraphics[width=0.95\columnwidth]{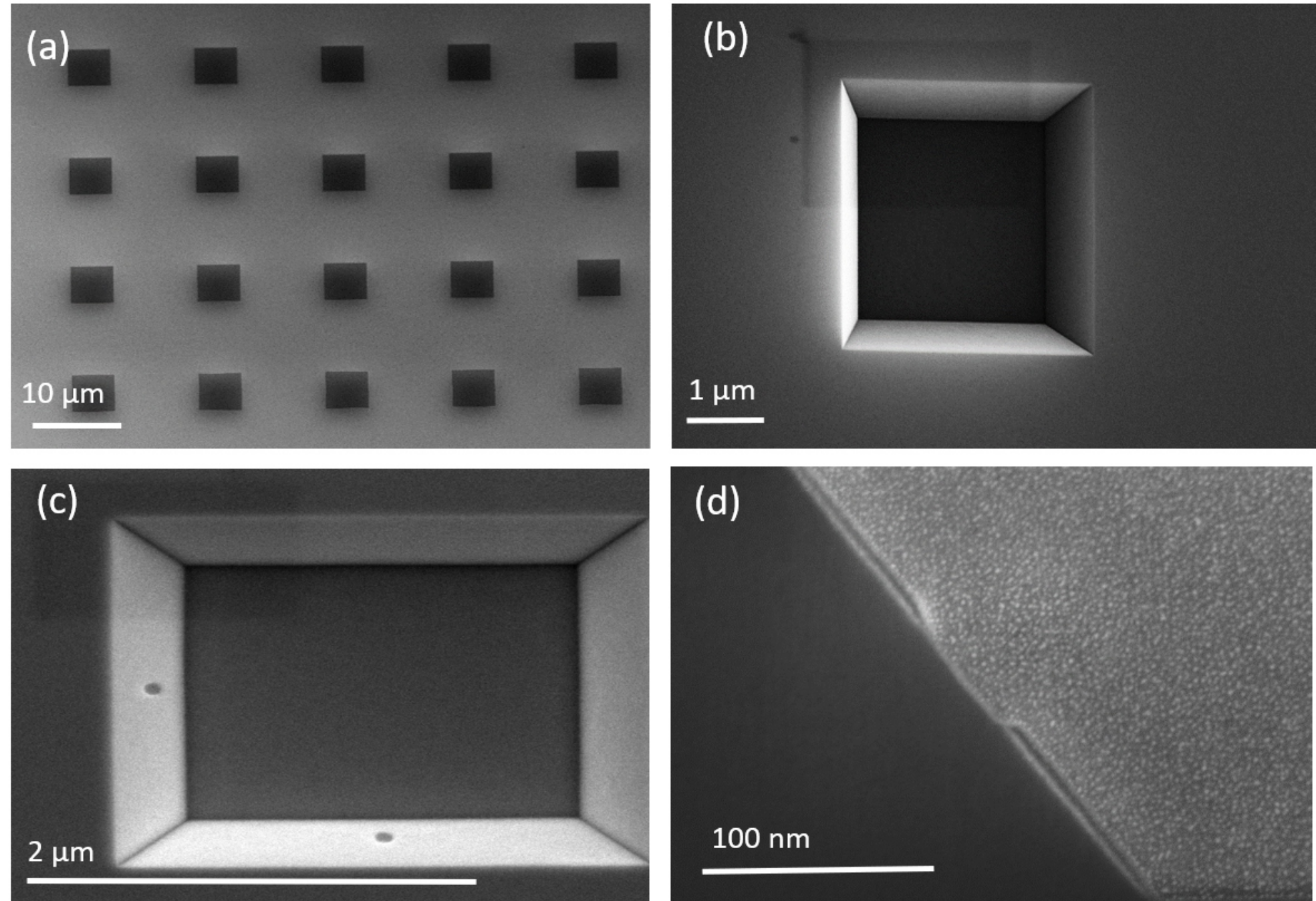}
	\caption{(a) SEM micrograph of 3 $\mu$m wide squares in a 20 nm SiO$_2$ layer, which has been treated by TMAH before to obtain Si (111) facets into Si (100). (b) 300\,nm deep square-shaped troughs with the Si(111) facets. (c) Etched hole after resist removal on Si (111) facet.  (d) Cross-section cut by focused ion beam of the oxidized square with PMMA resist (black) and the Pt metal (grey). The hole is etched into the SiO$_2$ layer.}
	\label{fig:SEM-3-step}
\end{figure}

Here, we provide a complete overview of the 3-step lithography process of etching holes into the tilted Si (111) facets. The images are taken by a scanning electron microscope (SEM). As depicted in Figure~\ref{fig:SEM-3-step} (a), several sets of 3\,$\mu$m wide squares with a pitch of 10\,$\mu$m are defined on the substrate surface by electron beam lithography. The SiO$_2$ cover of the substrate is anisotropically etched by reactive ion etching using CHF$_3$ and oxygen revealing the Si(100) surface. After resist removal the Si(100) surface is etched with tetramethyl ammonium hydroxide (TMAH) to form the 300\,nm deep square-shaped troughs with the Si(111) facets (cf. Figure~\ref{fig:SEM-3-step} (b)). Before the definition of the growth hole the oxide on substrate surface is removed completely with buffered HF followed by a thermal re-oxidation. The 80-nm-wide holes on the Si(111) side facets are defined by electron beam lithography and HF wet etching (cf. Figure~\ref{fig:SEM-3-step} (c)). In Figure~\ref{fig:SEM-3-step} (d) a cross-sectional cut of a 80\,nm wide hole on a Si (111) facet is depicted.

\subsection{Cross-sectional view of etched square} 
\begin{figure}[h!]
	\centering
	\includegraphics[width=0.95\linewidth]{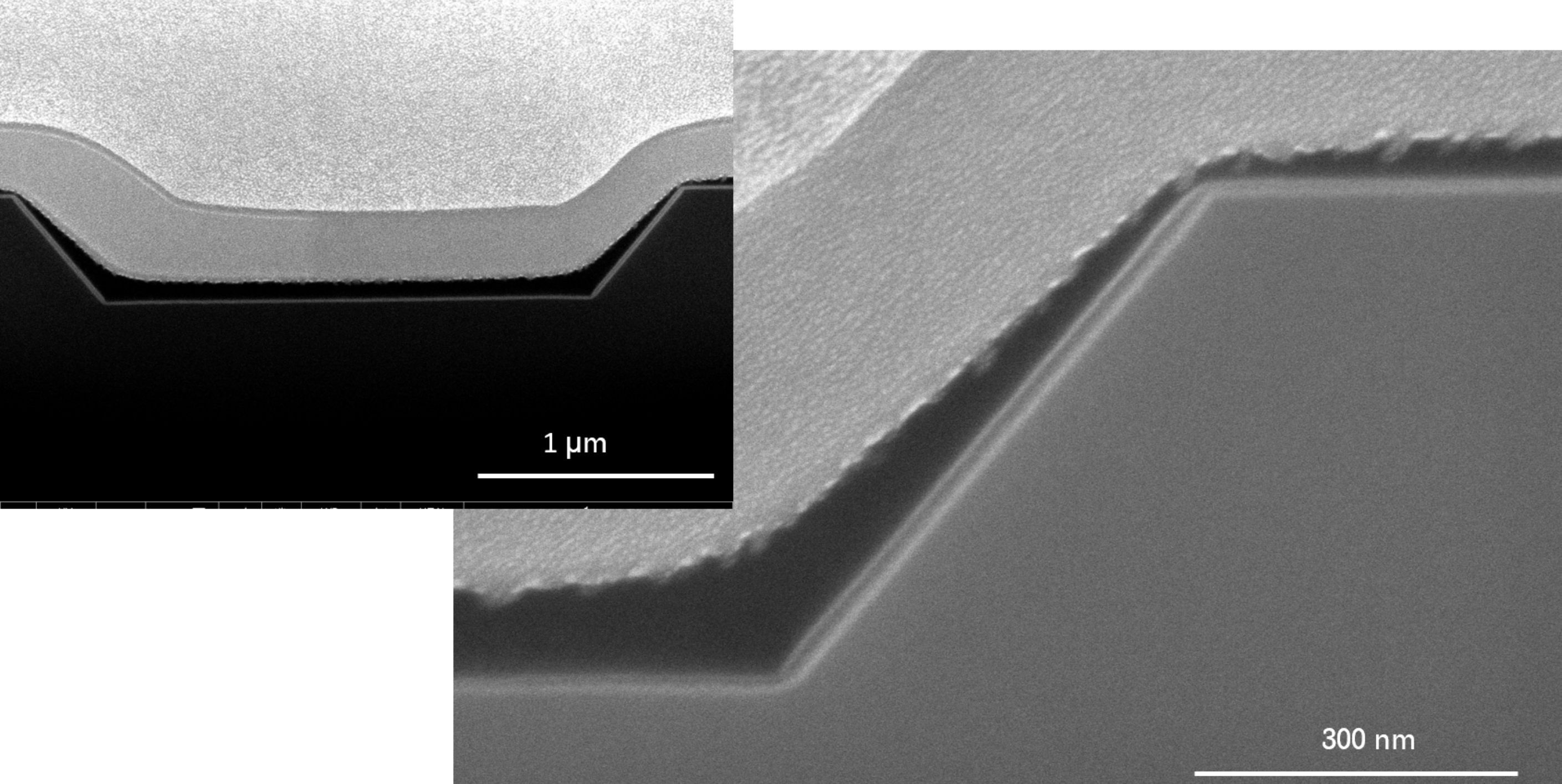}
	\caption{The complete profile of the PMMA resist in black and Pt sputtered in grey on top of the samples for protection and grey layer below the resist one finds the SiO$_2$ layer. The resist gets thinner towards the outer edges when compared to the inner part of the square. The thickness of the oxide layer varies on different Si facets, i.e. 23\,nm on the Si (111) facet and 16\,nm on the Si(100) facet.}
	\label{fig:resist-profile}
\end{figure}

Figure~\ref{fig:resist-profile} shows a focused ion beam (FIB) cross-sectional view of one of an etched square covered by a PMMA resist layer. The Si surface is covered by a SiO$_2$ layer. One finds that the resist becomes thinner towards the outer edges when compared to the inner part of the square.

\subsection{Processing details for substrate preparation and growth}

\begin{figure}[h!]
	\centering
	\includegraphics[width=0.95\columnwidth]{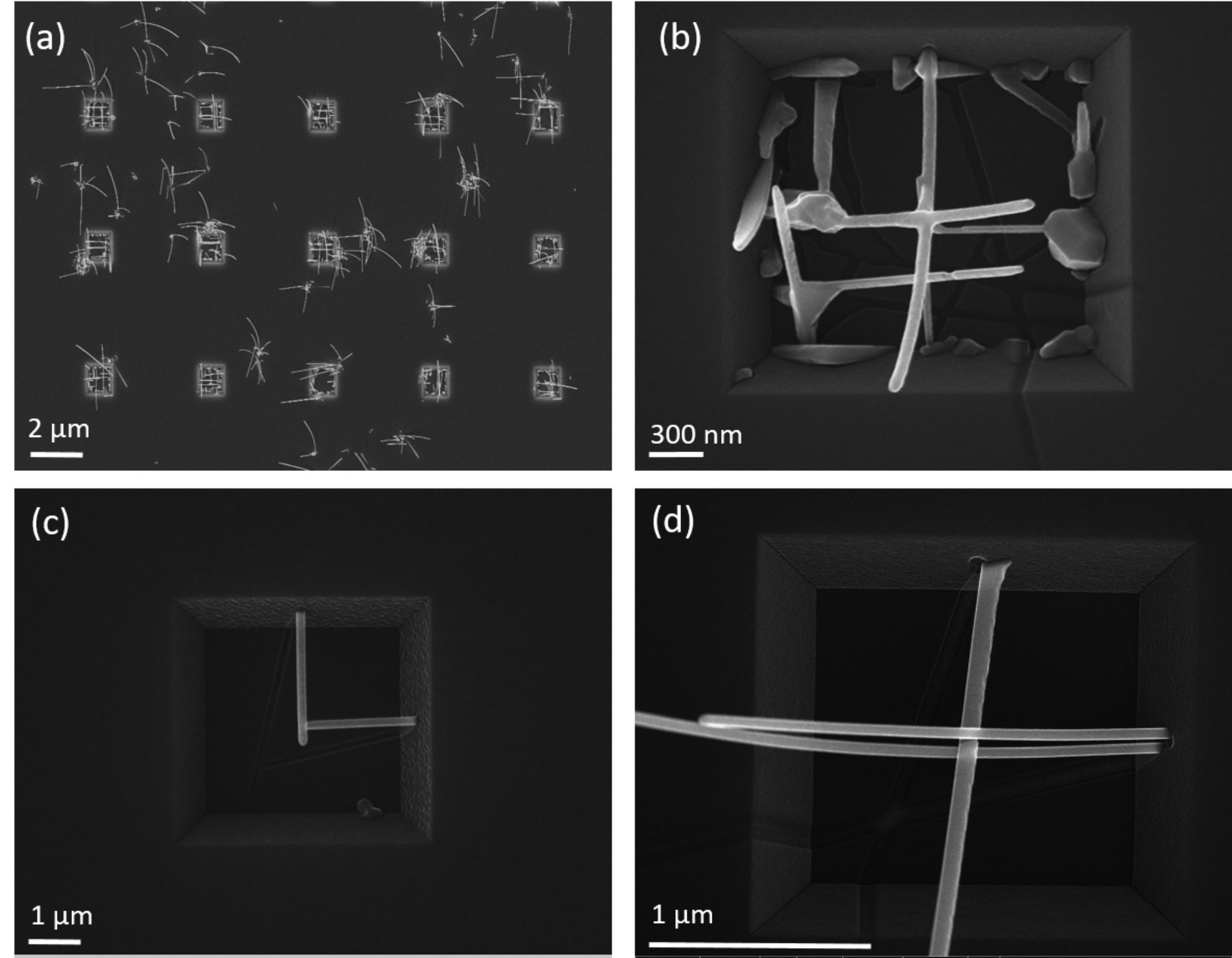}
	\caption{ (a) Hard mask (SiO$_2$) etched away due to PMMA stripping during HF treatment due to insufficient sticking of PMMA to SiO$_2$. (b) Thinner oxide in the bottom edges of the Si(111) causes nucleation during the growth i.e. parasitic growth. (c) Low arsenic fluxes causes stunted nanowire growth. (d) Larger electron beam lithography dose for the holes leading to the growth of two nanowires from one hole.}
	\label{fig:sae-probs}
\end{figure}

During the development of the sample preparation various issues occurred.
Resist sticking problems (cf. Figure~\ref{fig:sae-probs} (a)) have been tackled by bombarding the SiO$_2$ surface with 20 sec of O$_2$ (\textit{giga batch} tool) at 200 W, 300 sscm and then treating it with HMDS at 130$^{\circ}$C and finally spinning the resist. Oxide irregular growth on different Si facets has been tackled by growing a sufficiently thick oxide on Si (100) and Si (111) facets, i.e. 18\,nm and 25\,nm, respectively. After adjusting the RIE and HF etching parameters the parasitic growth shown in Figure~\ref{fig:sae-probs} (b) could be suppressed. In order to prevent stunted nanowire growth (cf. Figure~ \ref{fig:sae-probs} (c)), a moderately high As flux of $3.5 \times {10^{-5}}$ torr was used during the growth. In case that the dose for electron beam lithography is too high the holes in the SiO$_2$ layer are getting too large. This results in the growth of two nanowires per hole, as can be seen in Figure~\ref{fig:sae-probs} (d). In order to prevent this, the electron beam lithography parameters have been varied to find the right doses to get 1 for the  nanowire/hole diameter ratio.

\subsection{Randomly grown nanowires}

\begin{figure}[h!]
	\centering
	\includegraphics[width=0.95\columnwidth]{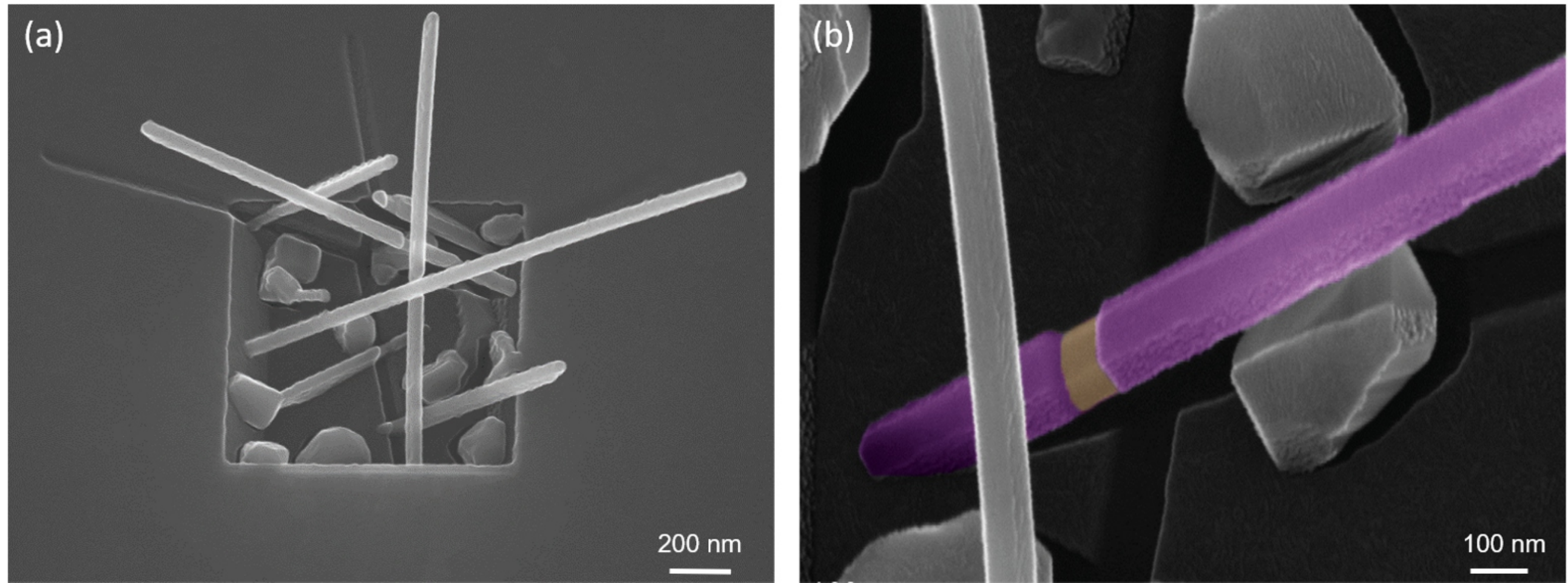}
	\caption{ (a) Random nanowire growth from every side and parasitic growth due to bigger nucleation sites. Before growth the Si(111) facets were treated with H$_2$O$_2$ to create pin holes. (b) Detail showing a randomly grown nanowire with Nb shell (pink) leaving a weak-link on the nanowire due to shadowing during the metal deposition.}
	\label{fig:random1}
\end{figure}

As a proof of concept for growing Josephson junctions with a weak-link in-situ InAs nanowires were also grown randomly on Si(111) facets. The processing is less elaborate, since it does not require the preparation of holes on the Si(111) facets. In case of random growth the TMAH etched squares have been treated with H$_2$O$_2$ for 120 seconds to create pin holes in the oxide. At these pin holes the nucleation takes place during the nanowire growth. Since this is an uncontrolled process the weak-link can be formed anywhere depending on the random position of the shadowing wire (cf. Figure~\ref{fig:random1}) 

\section{Transmission electron microscopy}

\begin{figure*}[h!]
	\centering
	\includegraphics[width=0.85\textwidth]{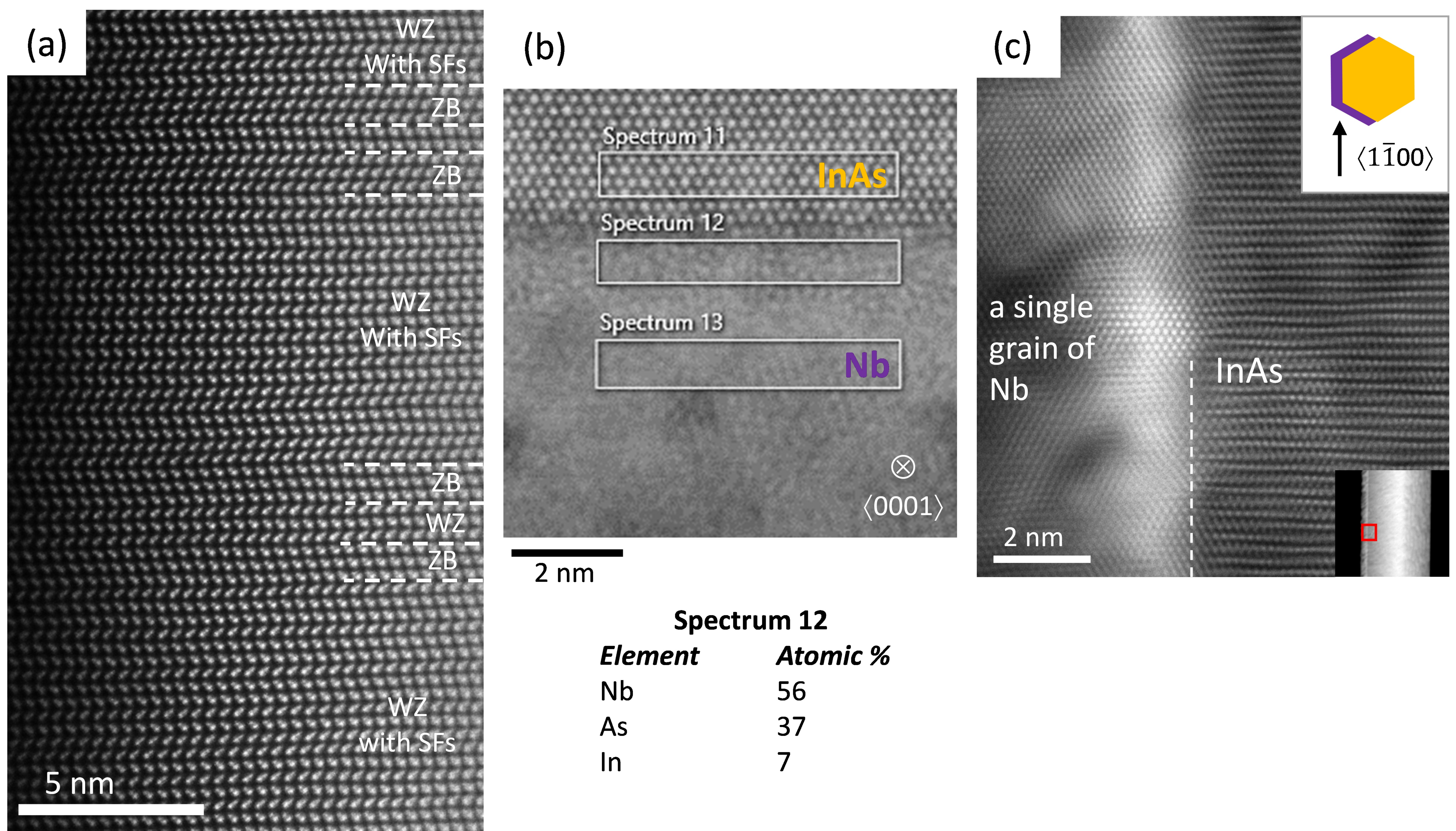}
	\caption{
		(a) Annular dark field (ADF) STEM image of a nanowire grown using the selective growth technique, showing the polytypic crystal structure. (b) Method of ascertaining composition of the $\sim$1\,nm thin amorphous layer, along with an example of the values extracted. (c) ADF STEM image of a large Nb grain oriented such that  [110]Nb$||$[0001]InAs in the axial direction. The red square in the bottom right inset indicates the area of acquisition and the top right inset shows a schematic indicating projections in the $<\!1\bar{1}00\!>$ viewing direction. The white broken line indicates the \{$11\bar{2}0$\} nanowire facet edge. However, note that part of the Nb shell is still viewed on the nanowire side along the projection, due to the half-shell growth on multiple facets around the nanowire as shown by the inset schematic. 
	}
	\label{fig:TEM-Supp-1}
\end{figure*}

As seen in Figure~\ref{fig:TEM-Supp-1} (c) and similar to the previous observation related to Al half-shells,\cite{Zellenkens20a} a single Nb grain (which is larger than $\sim$15\,nm) can grow beyond the length of a typical polytypic region (assuming that the amorphisation takes place later through solid diffusion, as it is the more likely case). This extended growth of Nb beyond the polytypic segment size could be due to the nucleation-and-expansion type growth process of metal grains or Nb growth being driven by factors other than those related to surface and interface.\cite{Kanne20} However, the possibility of the rough micro-facets formed by ZB inclusions leading to more nucleations (and hence smaller grain size) cannot be discounted.\cite{Ghalamestani12}

%
\end{document}